# One-Dimensional Luttinger Liquids in a Two-Dimensional Moiré Lattice

Pengjie Wang[1,#], Guo Yu[1,2,#], Yves H. Kwan[3,#], Yanyu Jia[1], Shiming Lei[4,5], Sebastian Klemenz[4,6], F. Alexandre Cevallos[4], Ratnadwip Singha[4], Trithep Devakul[7], Kenji Watanabe[8], Takashi Taniguchi[9], Shivaji L. Sondhi[1,3], Robert J. Cava[4], Leslie M. Schoop[4], Siddharth A. Parameswaran[3], Sanfeng Wu[1,*]

[1] Department of Physics, Princeton University, Princeton, New Jersey 08544, USA
[2] Department of Electrical Engineering, Princeton University, Princeton, New Jersey 08544, USA
[3] Rudolf Peierls Centre for Theoretical Physics, University of Oxford, Oxford OX1 3PU, United Kingdom
[4] Department of Chemistry, Princeton University, Princeton, New Jersey 08544, USA
[5] Department of Physics and Astronomy, Rice University, Houston, Texas 77005, USA
[6] Fraunhofer Research Institution for Materials Recycling and Resource Strategies IWKS, Hanau D-63457, Germany
[7] Department of Physics, Massachusetts Institute of Technology, Cambridge, Massachusetts 02139, USA
[8] Research Center for Functional Materials, National Institute for Materials Science, 1-1 Namiki, Tsukuba 305-0044, Japan
[9] International Center for Materials Nanoarchitectonics, National Institute for Materials Science, 1-1 Namiki, Tsukuba 305-0044, Japan

[#] These authors contributed equally to this work
[*] Email: sanfengw@princeton.edu

## Abstract

The Luttinger liquid (LL) model of one-dimensional (1D) electronic systems provides a powerful tool for understanding strongly correlated physics including phenomena such as spin-charge separation[1]. Substantial theoretical efforts have attempted to extend the LL phenomenology to two dimensions (2D), especially in models of closely packed arrays of 1D quantum wires[2–19], each being described as a LL. Such coupled-wire models have been successfully used to construct 2D anisotropic non-Fermi liquids[2–6], quantum Hall states[7–14], topological phases[15–17], and quantum spin liquids[18,19]. However, an experimental demonstration of high-quality arrays of 1D LLs suitable for realizing these models remains absent. Here we report the experimental realization of 2D arrays of 1D LLs with crystalline quality in a moiré superlattice made of twisted bilayer tungsten ditelluride (tWTe$_2$). Originating from the anisotropic lattice of the monolayer, the moiré pattern of tWTe$_2$ hosts identical, parallel 1D electronic channels, separated by a fixed nanoscale distance, which is tunable by the interlayer twist angle. At a twist angle of ~ 5 degrees, we find that hole-doped tWTe$_2$ exhibits exceptionally large transport anisotropy with a resistance ratio of ~ 1000 between two orthogonal in-plane directions. The across-wire conductance exhibits power-law scaling behaviors, consistent with the formation of a 2D anisotropic phase that resembles an array of LLs. Our results open the door for realizing a variety of correlated and topological quantum phases based on coupled-wire models and LL physics.

## Main

In various coupled-wire models[2–19], 1D quantum wires are placed in parallel with each other at an exactly fixed nanoscale distance, producing a 2D or 3D periodic system. In 2D, such perfectly arranged wires can in principle realize a strongly anisotropic non-Fermi liquid phase that resembles a LL[2–6]. When a perpendicular magnetic field is applied, novel quantum Hall states[7–14] may also develop in such an array without the presence of a free 2D electron gas. This highly anisotropic setting is qualitatively different from conventional isotropic 2D electron systems. Experimentally realizing these interesting coupled-wire constructions is challenging as they require a large number of identical nanowires to be strictly arranged in a crystalline array at both nano and microscopic scales. A route to overcome these difficulties is to use moiré superlattices of a twisted bilayer stack of an anisotropic 2D crystal. Indeed, it has been proposed that twisted 2D crystals with a rectangular unit cell, such as GeSe[20], create 1D flat bands. Another excellent choice is tWTe$_2$, as its monolayer's unit cell is an elongated rectangle. In this work, we uncover the potential of tWTe$_2$ for creating the desired high-quality arrays of 1D wires that can expand the LL physics to 2D.

**tWTe$_2$ Moiré Lattices and Device Design**

Monolayer WTe$_2$ consists of three atomic layers (Te-W-Te) in a sandwich structure, where the W atoms are organized in 1D zigzag chains[21] (Fig. 1a). The tWTe$_2$ hence has six atomic layers with a complicated moiré pattern. To better illustrate the moiré lattice of small-angle tWTe$_2$, we present the superlattice of only the W layers and that of only the Te layers separately in Fig. 1b and c. The Te pattern develops a triangular superlattice viewed from the top, while the W moiré pattern develops 1D stripes, reflecting the underlying anisotropy of the monolayer. The bright stripes in Fig. 1b indicate regions where the W atoms from two layers are optimally aligned vertically (AA stripes), while in the dark stripes they are optimally misaligned (AB stripes). The distance $d$ between neighboring AA stripes depends on the twisted angle $\theta$, $d = a/[2\sin(\theta/2)]$, for small $\theta$ (Fig. 1d); here $a$ is the length of monolayer's rectangular unit cell. In Fig. 1e and Extended Data Fig. 1, we experimentally visualize this unique moiré structure of tWTe$_2$ using conductive atomic force microscopy (cAFM). Below we present transport studies of two devices with $\theta \sim 5°$ ($d \sim 7.2$ nm, device #1) and $\sim 6°$ ($d \sim 6.0$ nm, device #2).

Figure 1f illustrates the design of our devices for measuring the transport properties of tWTe$_2$. Similar to our previous reports[22,23], a thin hexagonal boron nitride (hBN) layer is inserted between the tWTe$_2$ and the palladium (Pd) electrodes, with selected areas etched in the thin hBN layer that expose the very ends of Pd for contacting the tWTe$_2$ bulk. Such a device geometry restricts the contact area to be small in the 2D bulk and eliminates transport contributions of conducting edges or any extra monolayer regions next to the stack. The tWTe$_2$ is fully encapsulated with hBN/graphite stacks from both top and bottom, which also act as the electrostatic gates. An optical image of a typical device is shown in Fig. 1g. We fabricate multiple contacts in a ring structure, key to investigating transport anisotropy. Details about the fabrication procedure are described in Methods and Extended Data Fig. 2.

**Exceptionally Large Transport Anisotropy**

We first examine transport anisotropy in the tWTe$_2$ devices. Figure 1i plots the four-probe resistance $R_{xx}$ as a function of gate-induced doping $n_g$, taken from device #1 at 1.8 K with the contact configurations ($R_{hard}$ and $R_{easy}$) shown in Fig. 1h. Here $n_g \equiv \varepsilon_r\varepsilon_0(V_{tg}/d_{tg}+V_{bg}/d_{bg})/e$, where $d_{tg}$ ($d_{bg}$) is the thickness of hBN dielectric layers for the top (bottom) gate; $V_{tg}$ ($V_{bg}$) is the top (bottom) gate voltage; $e$, $\varepsilon_0$ and $\varepsilon_r$ are the elementary charge, vacuum permittivity and relative dielectric constant of hBN, respectively. The current is applied along two orthogonal directions in the atomic plane for measuring $R_{hard}$ and $R_{easy}$, respectively. The choice of the easy and hard directions was made by examining two-probe resistances taken between all neighboring electrodes (Extended Data Fig. 3) and other configurations along and across moiré stripes (Extended Data Fig. 4), where substantial anisotropy already appears. The four-probe measurement presented here eliminates the contact resistance and provides a better evaluation of the anisotropy. As seen in Fig.

1i, while $R_{hard}$ and $R_{easy}$ are close when tWTe$_2$ is electron-doped, but the two curves dramatically deviate from each other when the doping is changed to the hole-side. Similar behaviors were also observed in device #2 (Fig. 1j & Extended Data Fig. 5). Figure 1k plots the anisotropy ratio, $\beta_{4p} \equiv R_{hard}/R_{easy}$, which approaches ~ 1,000 in the hole-doped region, an exceptionally large value. The intrinsic resistivity anisotropy may be estimated as large as ~ 50 based on an electrostatic simulation considering the effect of measurement geometry (Extended Data Fig. 6)

Both varying the doping to the electron side or warming up the sample (Extended Data Fig. 5 and 7) strongly suppress $\beta_{4p}$ to near unity, suggesting that the large transport anisotropy is an intrinsic low-temperature property of hole carriers. Indeed, the correlation between the anisotropy and the hole doping is clearly seen in Fig. 2a, where we map out $\beta_{4p}$ as a function of both $V_{tg}$ and $V_{bg}$ for device #1 at 1.8 K (see also Extended Data Fig. 7). The transition from a nearly isotropic phase to a highly anisotropic phase occurs when the hole carriers become dominant, regardless of the electric displacement field.

**Conductance Power Laws**

The strongly anisotropic phase of tWTe$_2$ exhibits robust power-law and scaling behaviors in the across-wire transport (Fig. 2b), where currents flow perpendicular to the moiré stripes (i.e., the wires). Figure 2c plots the measured across-wire conductance $G$ at a selected gate voltage in the hole-doped regime, with a two-probe configuration shown in Fig. 2b. As seen in the log-log plot, $G \propto T^\alpha$, for $T$ below ~ 30 K, with an exponent $\alpha$ ~ 0.98 (~ 1.14) for device #1 (#2) at the chosen gate voltages. To demonstrate scaling, we present differential conductance (d$I$/d$V$) measurements under varying both the d.c. source-drain bias ($V$) and $T$ (Fig. 2d). For small enough bias, d$I$/d$V$ develops plateaus, indicating that the conductance is controlled only by $T$ via the power law. At high bias, all curves taken at different $T$ merge together with a trend that can be well captured by the *same* power-law exponent $\alpha$, i.e., d$I$/d$V \propto V^\alpha$ (the dashed line). Indeed, in the scaled conductance plot (Fig. 2e), (d$I$/d$V$)/$T^\alpha$ v.s. $eV/k_BT$, all data points, taken in a parameter range wider than a decade in $T$ and three decades in $V$, collapse into a single curve[24,25]. A similar collapse can be found in device #2 (Fig. 2f & Extended Data Fig 8).

In Extended Data Fig. 9, we compare along-wire and across-wire conductance taken from device #1. More robust power laws are typically seen in the across-wire direction. We note that contact resistance plays an important role in the along-wire transport, as seen in nanotubes[24]. In practice, the moiré system in the contact region may be affected by distortions, strain, unintentional doping, and other interface effects. In contrast, across-wire resistance at low $T$ is dominated by the tWTe$_2$ bulk (see Extended Data Fig. 10), a much more uniform area.

**Gate-Tuned Anisotropy Crossover**

The power-law across-wire conductance is generally observed over a wide range of gate voltages for $T < 30$ K in the tWTe$_2$ samples, as illustrated in Fig. 3 and Extended Data Fig. 11-13. We extract the $n_g$ dependent $\alpha$ together with $\beta_{4p}$ for devices #1 and #2 respectively (Fig. 3a and e). In the hole-doped side, strong anisotropy occurs together with good power-law scaling characteristics, as shown by the collapse of the d$I$/d$V$ curves over a wide range of $T$ and $V$ in the scaled plots (Fig. 3b-d for device #1; Fig. 3f-h for #2). Although in both devices $\alpha$ is valued near unity on the hole side, the exact gate-dependent behavior differs, which could arise due to twist-angle dependent electronic structures or extrinsic effects such as disorders. Near charge neutrality or on the electron-doped regime, high-bias data deviates from the power-law trend (Extended Data Fig. 12 & 13). With electron doping, transport anisotropy is strongly suppressed, although zero-bias $G(T)$ still approximately follows a power law with a decreasing exponent down to near zero at high electron doping (Extended Data Fig. 11).

**Band Structure Modeling**

We further perform a continuum model analysis on tWTe$_2$ at the single-particle level (Fig. 4). The modeling is challenging as even at the monolayer level, topology[26–29], correlations[22,23,30,31] and spin-orbit coupling[26] are all present. We start with a density functional theory (DFT) calculation on the monolayer, yielding valence band maximum at Γ flanked by two conduction band minima at the wavevector $\pm q_c$ (Fig. 4a). DFT calculations for untwisted but shifted bilayers are used to extract effective interlayer couplings (Fig. 4b & c), which enter the continuum model for obtaining the tWTe$_2$ structure[32]. The resulted twisted bands arising from one conduction valley are shown in Fig. 4d, where a pair of highly anisotropic bands indeed develop in the hole regime, in contrast to the electron regime where no substantial anisotropy is seen. Figure 4f illustrates the corresponding quasi-1D hole Fermi surface with the corresponding real space wavefunctions that coincide with the moiré stripes (Fig. 4e). In contrast to the $\pm q_c$ valleys, the moiré reconstruction of the valence bands at Γ is much less pronounced in this simplified model and develops no large anisotropy (Methods). Note that while our simple analysis here does capture the emergence of quasi-1D bands, a comprehensive modeling would necessarily require future efforts involving large-scale DFT calculations, lattice reconstructions and interaction effects.

**The Luttinger Liquid Interpretation**

The large resistance anisotropy and contrasting $T$ dependence of $R_{\text{hard}}$ and $R_{\text{easy}}$ (Extended Data Fig. 5) indicate that transport is qualitatively different between across- (insulating) and along-wire (metallic) directions. The power-law behavior itself is inconsistent with the formation of an ordinary band or Mott insulator. The exponent $\alpha$ varies smoothly as a function of $n_g$, showing no obvious presence of a fully insulating state (Fig. 3), consistent with the absence of a gap in the modeling. For a 2D diffusive metal[33,34], the "tunneling anomaly" owing to the relaxation of injected charges at the contact may lead to a conductance power law depending on $T$ or $V$. However, this cannot give distinct transport exponents in different directions and cannot account for our observation along the hard direction where the resistance is dominated by the tWTe$_2$ bulk rather than contact effects. For disordered quasi-1D systems, calculations have shown that variable range hopping transport may produce an apparent power-law behavior[35], i.e., $G \propto T^\alpha$ for $eV \ll k_B T$ and $dI/dV \propto V^\beta$ for $k_B T \ll eV$, where $\alpha$ and $\beta$ are two generally unequal exponents that are independently controlled by microscopic details including disorder. This is however in sharp contrast to our observation in tWTe$_2$, where the power laws in $T$ and $V$ are controlled by the same exponent, i.e., $\alpha = \beta$. This single exponent scaling behavior is robustly observed over a wide hole doping range (Fig. 3, Extended Data Fig. 12 & 13), where $\alpha$ has been tuned, and in samples with varied twist angles (Devices #1 & #2). These observations provide strong evidence that the single-exponent power-law behavior is generic to the anisotropic phase of tWTe$_2$. Any explanation that requires fine-tuning of parameters to achieve the condition of $\alpha = \beta$ is unlikely to be feasible.

A natural explanation is the emergent LL physics. The characteristic feature of a 1D LL in transport is indeed the *single exponent* power-law dependence of its conductance, i.e., $\alpha = \beta$. The power-law transport of LL physics has been observed in several 1D systems, such as nanotubes[24,25,36,37], engineered 1D structures[37–39], edge modes[40–42], polymers[43], and self-organized gold wires[44]. However, extending LL physics from a single 1D wire to a 2D system is of fundamental interest yet challenging. Proposals to do so based on 2D arrays of 1D wires have been put forward[2–6], but as far as we know, the intriguing concept of a 2D anisotropic phase that mimics a LL has so far not been established in real materials.

Our observations on the hole side of tWTe$_2$ are well-consistent with the generic LL expectations. We hence propose that the anisotropic phase arises due to the formation of a 2D array of 1D LLs induced by the moiré superlattice. Understanding the moiré-induced LL behaviors in tWTe$_2$ requires proper consideration of electron interactions and transport mechanisms. In a quasi-1D system, while the early calculations[45] of the across-wire transport exponent between parallel LLs suggested $\alpha = 2\eta$, where $\eta$ (which vanishes without interactions) is the Fermi surface exponent for an individual wire determined by the LL parameter $K$, the more recent consensus is instead $\alpha = 2\eta - 1$, where the extra -1 arises from the fact that hopping can occur anywhere along the wires[46]. The relation applies when single-particle hopping is the dominant conduction process and $T$ is much larger than the 1D-to-2D crossover temperature $T^* \sim t_\perp (t_\perp/t_{//})^{\eta/(1-\eta)}$, where $t_\perp$ ($t_{\parallel}$) is the inter- (intra-) wire hopping[46]. From the energy scales of quasi-1D bands obtained in Fig. 4d, we may estimate $t_\perp \sim 5$ meV and $t_\parallel \sim 50$ meV respectively. If this applies, for certain hole doping of tWTe$_2$ the across-wire conductance exhibits a power-law exponent $\alpha \sim 1$, corresponding to an effective $\eta \sim 1$, near the marginal boundary above which the single-particle process is no longer relevant and two-particle processes may be important. Assuming spin degeneracy and $\eta = (K+1/K-2)/4$,[46] we obtain an effective intrawire $K \sim 0.17$ for $\alpha \sim 1$. The strong intrawire interaction is consistent with the experimental fact that the deviation from the power law is absent down to at least 1.8 K. This remarkably stable LL behavior in the anisotropic 2D system calls for careful consideration of the interaction-driven phases in tWTe$_2$, especially the interwire interactions given the nanoscale wire spacing. Considering interwire interactions, the transport exponent then depends on a stiffness function $\kappa(q_\perp)$ instead of a single intrawire parameter $K$, where $q_\perp$ is the momentum perpendicular to the wires[3–6]. We note that further experimental and theoretical explorations are necessary to examine the exact connection between the measured power laws to interactions in the system, which is critical to evaluate the enticing possibility of a sliding LL phase and a host of competing orders descending from it[2–6].

**Summary**

We demonstrate a novel tunable platform based on tWTe$_2$ stacks for studying high-quality 2D arrays of 1D electronic structures in a crystalline superlattice. We interpret the results based on the formation of a 2D anisotropic non-Fermi liquid phase that resembles a LL. An exciting direction is to search for novel quantum Hall states with an applied magnetic field[7–13]. The physics of spin-charge separation[1,37,38], naturally expected in LLs, is another interesting direction to pursue. Experimental searches for evidence of spin-charge separation in a 2D WTe$_2$ system could provide important opportunities for studying new regimes in strongly correlated quantum phases.

**Acknowledgements**

We acknowledge discussions with N. P. Ong, A. Yazdani, B. A. Bernevig, F.H.L. Essler, B. Lian, C. Kane, K. Yang, A. J. Uzan, Y. Werman, and J. Zhang. We especially thank S. H. Simon for discussions and for pointing out ref.[47] to us. This research was supported by NSF through a CAREER award to S.W. (DMR-1942942) and the Princeton University Materials Research Science and Engineering Center (DMR-2011750) through support to S.W., R.J.C, and L.M.S. Device characterization and data analysis were partially supported by ONR through a Young Investigator Award (N00014-21-1-2804) to S.W. S.W. and L.M.S. acknowledge the support from Eric and Wendy Schmidt Transformative Technology Fund at Princeton. Early measurements were performed at the National High Magnetic Field Laboratory, which is supported by NSF Cooperative Agreement No. DMR-1644779 and the State of Florida. K.W. and T.T. acknowledge support from the Elemental Strategy Initiative conducted by the MEXT, Japan, Grant Number JPMXP0112101001, JSPS KAKENHI Grant Number JP20H00354 and the CREST(JPMJCR15F3), JST. L.M.S. acknowledges the support from the Gordon and Betty Moore Foundation's EPIQS initiative through Grant GBMF9064 to L.M.S, the David and Lucile Packard Foundation, the Sloan Foundation, and Princeton's catalysis initiative (PCI). Y.H.K. and S.A.P. acknowledge support from the European Research Council under the European Union Horizon 2020 Research and Innovation Programme via Grant Agreement No. 804213-TMCS. S.L.S was supported by the Gordon and Betty Moore Foundation through Grant GBMF8685 towards the Princeton theory program and by a Leverhulme International Professorship at Oxford.

**Author Contributions**

S.W. conceived and designed the project. P.W. and G.Y. fabricated devices, performed measurement and analyzed data, assisted by Y.J. and supervised by S.W. Y.H.K, T.D., S.L.S., and S.A.P. performed theoretical modeling. F.A.C., R.J.C., S.L., S.K., R.S., and L.M.S. grew and characterized bulk $WTe_2$ crystals. K.W. and T.T. provided hBN crystals. S.W., P.W., S.A.P., and Y.H.K. wrote the paper with inputs from all authors.

**Competing Interests**

The authors declare no competing financial interests.

**Data Availability**

The data that support the findings of this study are available from the corresponding author upon reasonable request.

**Methods**

**Sample Fabrication**

We followed the $WTe_2$ crystal growth, exfoliation, and device fabrication procedures detailed in ref.[21–23,48]. The stack of $tWTe_2$ was obtained by the "tear and stack" technique[49,50], in which we used the hBN layer to tear and pick up part of a monolayer $WTe_2$ flake, followed with rotating the rest of the flake counterclockwise by a chosen angle $\theta$, and then stacking the two monolayer $WTe_2$ pieces together. We summarized a step-by-step fabrication process in Extended Data Fig. 2. The device for cAFM measurements consists of a stack of few-layer $hBN/tWTe_2/hBN$ on top of a layer of Pd thin film.

**Conductive Atomic Force Microscopy (cAFM) Measurement**

The cAFM measurements[51] were performed at room temperature in a Bruker Dimension Icon AFM with dry nitrogen purged into the acoustic shield to eliminate the oxygen and water and reduce $WTe_2$ degradation. A humidity as low as < 0.1% and no more than 5% was kept during the measurement. A PF-TUNA module equipped with an in-situ current amplifier was used. The device was biased with -2.5 mV d.c. voltage and the d.c. current through the tip was recorded. The 2D conductance image was then captured and plotted as shown in Extended Data Fig. 1b and 1c. To better view the moiré pattern, we have applied a standard flatten process and filtered out the 60 Hz electronic noise to the AFM image.

**Transport Measurement**

The electrical measurements of our devices were performed in a cryostat (Quantum Design Dynacool) equipped with a superconducting magnet. Standard lock-in measurements were taken with a frequency of 3 ~ 78 Hz. The four-probe measurements were performed by supplying an a.c. current of ~ 3 nA. The two-probe differential conductance measurements were carried out with a small a.c. excitation of 50 or 100 μV together with a d.c. excitation up to ~ 400 mV. A current preamplifier (DL Instrument 1211) and a voltage preamplifier (DL Instrument 1201) were used to improve signals. Two Keithley 2450 were used to control the top and bottom gates.

## Estimating Intrinsic Anisotropy from Four-Probe Measurements

Following Ref[47], we discuss the impact that sample geometry and contact placement have on four-probe resistance measurements. In particular, we are interested in how an intrinsic sheet resistivity anisotropy $\beta_{bulk} = \rho_{xx}/\rho_{yy}$ translates to an observed four-probe anisotropy $\beta_{4p} = R_{xx}/R_{yy}$. We address this by considering the electric potential distribution over the sample in the classical limit. To simplify the problem, we assume that *i*) no current leaks out of the tWTe$_2$ sample boundary; *ii*) the sample is characterized by a spatially uniform resistivity tensor diag($\rho_0\beta_{bulk}, \rho_0$); *iii*) the external current source/sink distribution $f(x,y)$ is modeled as delta functions at the current contacts; and *iv*) the voltage (and unused) contacts do not dramatically change the physics. By combining the continuity equation, Ohm's law, and Faraday's law, we obtain an anisotropic Poisson equation for the scalar potential with derivative boundary conditions

$$\left[\frac{1}{\beta_{bulk}}\partial_x^2\phi(x,y) + \partial_y^2\phi(x,y)\right] = -\rho_0 f(x,y)$$

$$\left(\frac{1}{\beta_{bulk}}\partial_x\phi(x,y), \partial_y\phi(x,y)\right) \cdot \widehat{n_S}(x,y) = 0 \quad \text{for } (x,y) \in \partial S$$

where $S$ denotes the tWTe$_2$ region. From this, the measured resistance can be obtained once the voltage contact locations are prescribed.

We consider a simple caricature of the experimental setup by taking the sample to be a square of side length $L$ aligned with the principal axes of the resistivity tensor, i.e. the region $[-L/2,L/2] \times [-L/2,L/2]$. For a measurement of $R_{xx}$, the current contacts are placed at $(\pm L/4, 0)$ and the voltage contacts at $(\pm L/8, L/4)$. $R_{yy}$ is computed similarly, except the contacts are rotated by 90°. The results are shown in Extended Data Fig. 6 for different values of the $\beta_{bulk}$. Above some critical value, the measured anisotropy $\beta_{4p}$ grows exponentially as the square root of $\beta_{bulk}$.

## Scaling Formula and the Fitting Procedure

To fit the conductance data shown in Fig. 2, we derive a scaling formula for across-wire transport by following the procedure and assumptions in ref. 52. The resulting formula is:

$$\frac{1}{T^\alpha}\frac{dI}{dV} = A\left(\cosh\gamma x\left[\left(\frac{1+\alpha}{4}\right)^2 + \left(\frac{\gamma x}{2\pi}\right)^2\right]|\Gamma(z)|^4 \right.$$

$$+ \sinh\gamma x \frac{\gamma x}{2\pi^2}|\Gamma(z)|^4$$

$$\left. + \frac{\sinh\gamma x}{\gamma}\left[\left(\frac{1+\alpha}{4}\right)^2 + \left(\frac{\gamma x}{2\pi}\right)^2\right]\left(\frac{i\gamma}{\pi}|\Gamma(z)|^4\psi(z) + h.c.\right)\right)$$

where the dimensionless variable $x = \frac{eV}{2k_BT}$, $z = \frac{1+\alpha}{4} + \gamma\frac{ix}{2\pi}$, $\Gamma(z)$ is the gamma function, $\psi(z)$ is the digamma function, $\alpha$ is the power-law exponent, $\gamma$ is a constant introduced to account for the division of the source-drain voltage across the multiple wires in series, and $A$ is an overall coefficient. This formula assumes that the dominant transport mechanism between parallel LL channels is single-particle hopping[52]. Note that we have taken a derivate on $I(V)$ to obtain an expression for d$I$/d$V$. In the fitting procedure, we first assign the $\alpha$ and $1/\gamma$ to specific values and find the best fit by optimizing the parameter $A$; for each combination of $\alpha$ and $1/\gamma$, a root mean square error (RMSE) considering all data points is calculated. The RMSE as functions of $\alpha$ and $1/\gamma$ is plotted as the 2D color plot (see details in Extended Data Fig. 8 for both device #1 and #2). The best fits fall in the regions with minimized RMSE in the 2D plots. We note that our modeling of the tWTe$_2$ system is currently at a very early stage and hence new scaling formula that better describes tWTe$_2$ transport may be developed in the future with improved understanding of the system. However, we emphasize that the key experimental demonstration of the power law scaling here is based on the direct observations of the single exponent behavior, i.e., $G \propto T^\alpha$ at the low bias limit, d$I$/d$V \propto V^\alpha$ at the high bias limit (Fig. 2d), and the fact that all data collapse in the scaled conductance plot (Fig. 2e). These key features are independent of any fitting formula used for analysis.

**Density Functional Theory Calculations**

Density functional theory (DFT) calculations on untwisted systems were performed using the plane-wave pseudopotential code QUANTUM ESPRESSO[53]. For the band structure of monolayer WTe$_2$, we used fully relativistic optimized norm-conserving Vanderbilt pseudopotentials from PseudoDojo[54] and the PBE exchange-correlation functional[55] with a 10×6×1 $k$-grid. Van der Waals corrections were included via the semi-empirical framework of DFT-D3[56]. The plane wave cutoff was 80 Ry and Marzari-Vanderbilt smearing of width 0.01 Ry was used. We consider a cell of height 35 Å to eliminate the effect of periodic images. The fully relaxed monolayer geometry was obtained without spin-orbit coupling (SOC) and van der Waals corrections, with a force tolerance of $10^{-4}$ a.u. on each atom. The shifted untwisted bilayers were constructed by fixing the in-plane positions of the atoms in one of the layers to be displaced by $\boldsymbol{d} = (dx, dy)$, but letting the out-of-plane coordinates to freely relax. The resulting band structures, obtained without SOC, determine the effective interlayer couplings used in the continuum models[32]. The neglect of SOC in the bilayers is a good approximation since the band splitting is mostly due to interlayer interactions and is relatively insensitive to SOC. Furthermore, the bands remain spin-degenerate since inversion symmetry is retained for any $\boldsymbol{d}$. While lattice relaxation is not considered, we expect its effect would be important in real devices as it will deform the heterostructure by expanding the low-energy stackings at the expense of the high-energy regions.

**Continuum Model Calculations**

Moiré continuum models can be constructed using input from monolayer and untwisted bilayer DFT data. Doing this for twisted WTe$_2$ is challenging because *i*) the unit cell is rectangular, *ii*) the (multiple) bands near the Fermi energy are complicated, and *iii*) some of the relevant low-energy features generally lie away from high-symmetry points. This is to be contrasted with hexagonal-based systems such as graphene (Dirac cones at $K$) or other transitional metal dichalcogenides (parabolic bands with large band gaps at $\Gamma$ or $K$). To make progress, we take the small-angle limit and assume that the main features of monolayer WTe$_2$ can be treated separately (i.e., the valence band maximum and conduction band minima (valleys)). While these bands have significant energetic overlap, for small twist angles and smooth moiré potentials, the coupling between these bands will be suppressed.

The general approach is to first model an effective Bloch Hamiltonian *H(k,d)* that describes the untwisted bilayer for different in-plane shifts *d* (it is assumed here that the out-of-plane coordinates have been relaxed)[32]. This typically involves kinetic energy terms *E(k)* from the individual layers, as well as (layer-dependent) interlayer potential *V(k,d)* and hopping terms *Δ(k,d)*. If we take one band from each layer, the effective Hamiltonian is

$$H(k, d) = \begin{pmatrix} E(k) + V_b(k, d) & \Delta(k, d) \\ \Delta^*(k, d) & E(k) + V_t(k, d) \end{pmatrix}$$

Then the relation $\boldsymbol{d} \sim \theta \hat{z} \times \boldsymbol{r}$, valid for rigid twists, is used to convert the interlayer interactions into 'hopping' terms in the moiré reciprocal lattice. Note that depending on the momentum basepoint, the twist will also induce a relative difference between the kinetic terms in the two layers. In particular, the conduction valley continuum model is based at $q_c$ which is around 1/3 the distance to the BZ boundary in the *a*-direction. The resulting matrix (in combined band/layer/moiré reciprocal lattice vector space) is diagonalized with a large enough plane wave cutoff to ensure convergence in the energy window of interest.

We apply the approach above to obtain the reconstructed moiré bands from a conduction band valley. SOC is neglected (as in the bilayer DFT calculations) and spin degeneracy is assumed (inversion is only weakly broken for small twists). The kinetic terms are taken as effective mass parabolas to emulate the band extrema in the monolayer: $m_x = m_y = 0.38 m_e$ for the valleys. The interlayer interactions *V(d)* and *Δ(d)* are taken to be independent of momentum. This is justified if the relevant momentum region is small since these quantities are already finite at the basepoint momenta. The symmetries of the monolayer, namely inversion, mirror, and time-reversal, can be used to constrain the *d*-dependence of the interlayer quantities, which are fit to the energy shift and splitting of the bilayer DFT bands at the basepoint momenta. The results are shown in Fig. 4. We further note that while the conduction valleys show a pronounced anisotropy on the hole-side, the continuum bands at the valence maximum do not show obvious signs of appreciable quasi-1D behavior in our modeling. On one hand, carriers originated from the valence maximum may be localized in real devices. This may be evidenced by the fact that the transport of monolayer $WTe_2$ is known to exhibit less mobility for hole doping compared to electron doping. On the other hand, we emphasize that more complete modeling is necessary to the development of a comprehensive understanding of the $tWTe_2$ moiré system. For example, the continuum model parameters were extracted by assuming a rigid rotation between two monolayers. It is known however that lattice relaxation and corrugation effects can be significant for small twist-angle moiré heterostructures[57,58]. Electron interactions are also expected to play an important role in the system. While the general prediction here of moiré-induced quasi-1D features should be robust, to obtain a more comprehensive picture of the twisted band structure it is crucial to perform large-scale fully relaxed DFT calculations on $tWTe_2$ and to consider electron interactions and SOC properly.

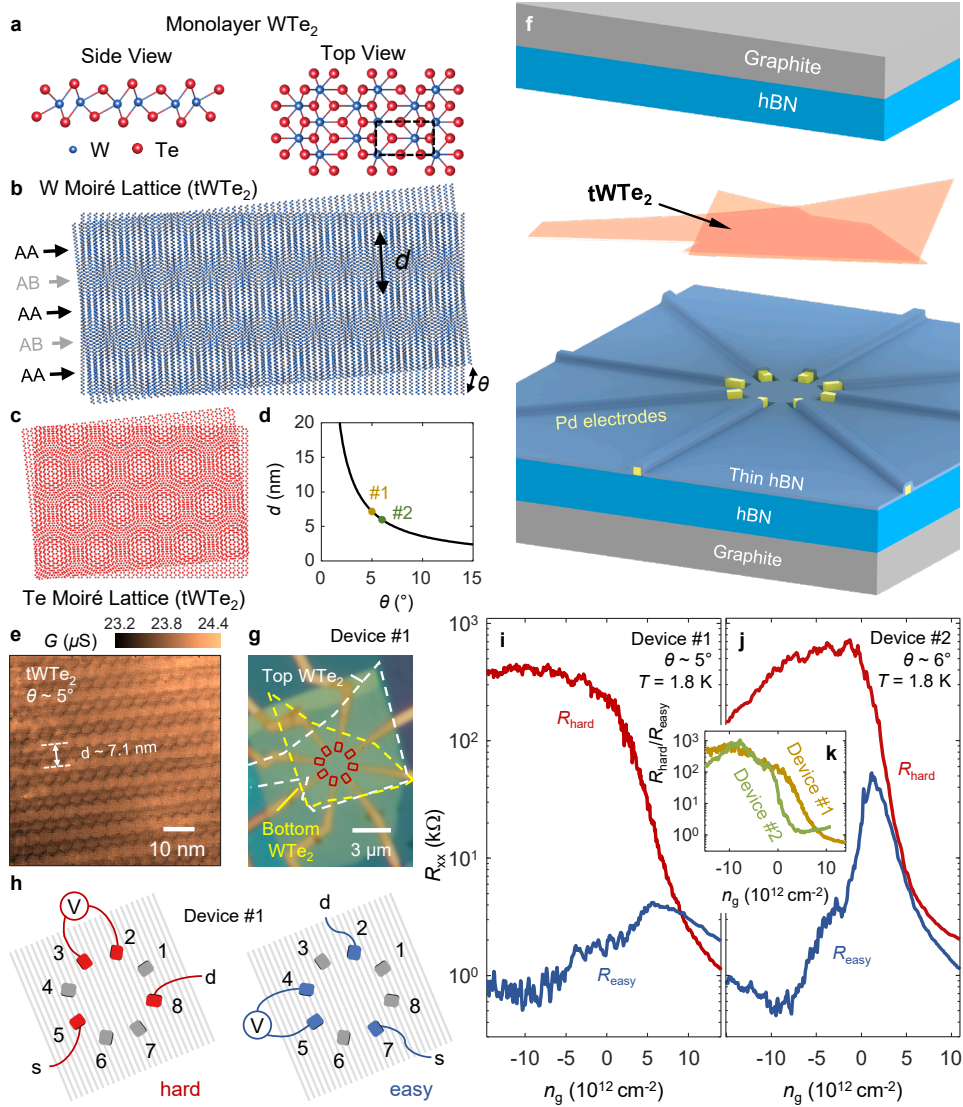

**Figure 1 | Small-angle tWTe$_2$ moiré lattices and large transport anisotropy. a,** Crystal structure of monolayer WTe$_2$ (left: side view; right: top view). The dashed rectangle indicates the unit cell. **b,** Moiré superlattice of W atoms only, showing 1D channels (AA and AB stripes). **c,** Moiré superlattice of the Te atoms, showing a 2D triangular pattern. **d,** Calculated distance $d$ between AA stripes shown in **b**, as a function of twist angle $\theta$. The yellow (green) point indicates the parameter realized in device #1 (#2). **e,** A cAFM image of the tWTe$_2$ moiré structure (see details of the measurement in Extended Data Fig. 1 and Methods). **f,** Cartoon illustration of our tWTe$_2$ device design, where a thin hBN layer with selectively etched areas is used to avoid electrical contact to non-tWTe$_2$ regions. **g,** An optical image of device #1. The dashed white (yellow) line highlights the top (bottom) monolayer WTe$_2$ and the red squares denote the contact regions. **h,** Cartoon illustration of the measurement configuration along hard and easy directions for device #1. The grey lines indicate the moiré stripes (not to scale). As an estimation, ~ 71 AA stripes exist between contacts 2 and 3. *s*: source; *d*: drain. **i,** Four-probe resistances measured with configurations shown in **h**, labeled as $R_{hard}$ and $R_{easy}$ respectively, as a function of $n_g$ for device #1 (cooldown #1) at 1.8 K. **j,** Similar four-probe resistances along easy and hard directions measured on device #2 (cooldown #1). **k,** The density-dependent anisotropy ratio, $R_{hard}/R_{easy}$, for the two devices, respectively.

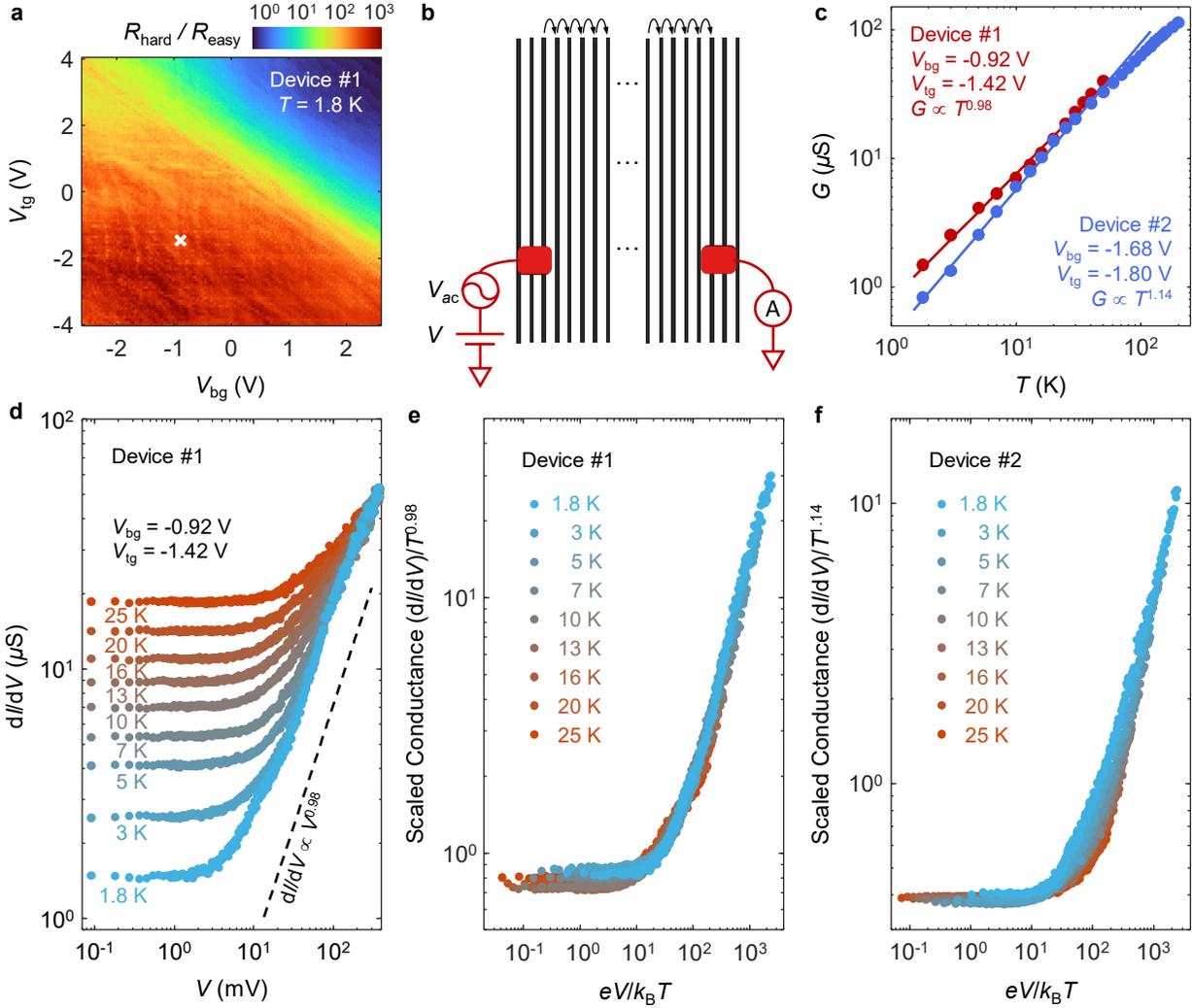

**Figure 2 | Luttinger liquid behaviors observed in the tWTe$_2$ devices. a,** Dual-gate dependent map of $R_{hard}/R_{easy}$ measured in device #1 (cooldown #1) at 1.8 K. **b,** Illustration of the measurement configuration for determining two-probe across-wire conductance $G$ and differential conductance $dI/dV$ used in **c-f**. **c,** across-wire conductance $G$ as a function of $T$, plotted in log-log scale for device #1 (red) and #2 (blue) at a selected gate parameter indicated by the cross in **a**. The solid lines are the power-law fittings. **d,** Across-wire differential conductance $dI/dV$ measured in device #1 as a function of d.c. bias $V$ at different $T$. **e,** The same data as **d**, but plotted as a scaled conductance $v.s.$ a scaled excitation. All data collapses to a single curve. **f,** the same scaled differential conductance plot taken in the hole-doped regime for device #2 (cooldown #1), with raw data included in Extended Data Fig. 8.

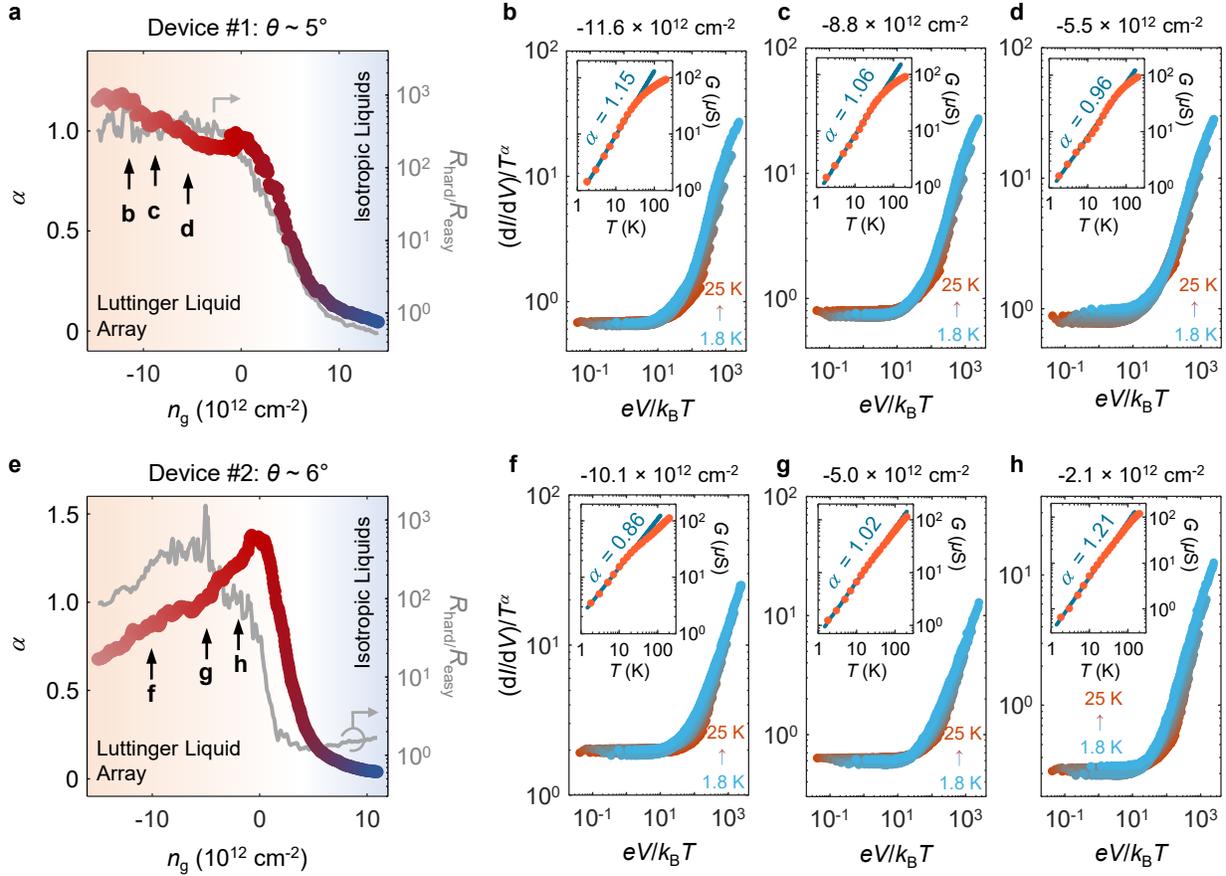

**Figure 3 | Gate-tuned power laws and anisotropy crossover. a,** Extracted power-law exponent $\alpha$ as a function of $n_g$ based on temperature dependent $G$ (raw data in Extended Data Fig. 11), measured in device #1 (cooldown #2). The grey curve plots the measured anisotropy ratio based on four-probe resistances along the easy and the hard direction in the same cooldown. **b-d,** The scaled differential conductance as a function of $eV/k_B T$ at different $n_g$, indicated in **a**. Insets show the corresponding $G(T)$, from which the corresponding $\alpha$ is extracted based on a power-law fit (solid line) to the low $T$ data. Raw conductance data are shown in Extended Data Fig. 12. **e-h,** The same plots for device #2 (cooldown #2) with raw conductance data shown in Extended Data Figs. 11 & 13.

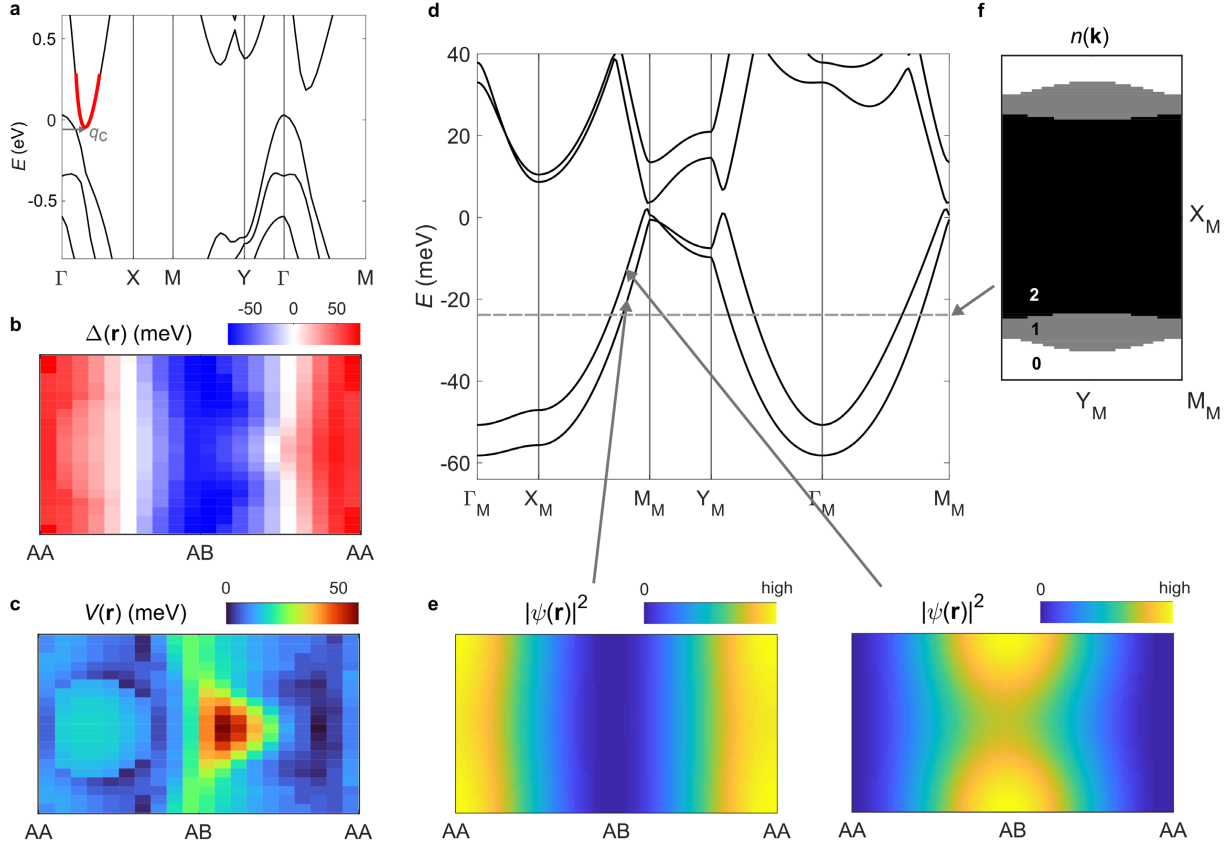

**Figure 4 | Theoretical modeling and the emergence of quasi-1D moiré bands at the single-particle level. a,** DFT band structure for monolayer WTe$_2$. Red shading highlights the conduction band valleys at $\pm q_c$, about which the continuum model analysis in **b-f** is performed. Results are shown for one of the valleys. Energies are measured relative to the Fermi energy at charge neutrality. **b & c,** Interlayer hopping and potential terms respectively, plotted in the moiré unit cell. These quantities are extracted from DFT calculations of untwisted bilayers with in-plane shift $d \sim \theta \hat{z} \times r$, valid for a rigidly twisted tWTe$_2$. AA and AB indicate the positions of W superlattice chains (see Fig. 1b). **d,** Continuum model band structure for a conduction band valley, plotted along a cut in the moiré BZ. Upon hole-doping, the system enters a highly anisotropic regime induced by the moiré physics. **e,** Representative Bloch wavefunctions in the quasi-1D regime plotted in the moiré unit cell. **f,** Illustrations of the quasi-1D open Fermi surfaces for moderate hole-doping, with the number of occupied quasi-1D bands indicated.

**Extended Data Figures**

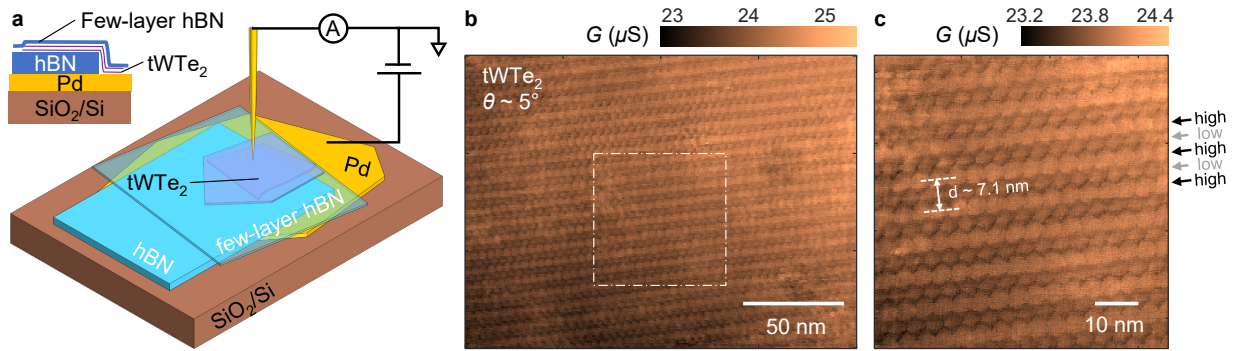

**Extended Data Fig. 1 | Conductive atomic force microscope (cAFM) measurements of tWTe$_2$ at room temperature. a,** Cartoon illustrations of the cAFM measurement on a few-layer hBN/tWTe$_2$/hBN stack on a Pd metal film pre-patterned on a SiO$_2$/Si chip. The inset illustrates the cross-section of the stack. A relatively thick (~ 39 nm) hBN was used to mitigate the roughness of the metal surface. **b,** A cAFM image taken from a $\theta \sim 5°$ tWTe$_2$ device, directly visualizing the moiré structure. The dashed-dot square locates a zoom-in scan, as shown in **c**. We comment on three aspects of the observations. (1) The measurement was taken at room temperature, where the transport shows no significant anisotropy. This is consistent with the fact the measured local conductance $G$ varies only by a small amount at different tip locations in the map. (2) As shown in **b** and **c**, the small variations already allow us to clearly image the underlying moiré structure. (3) Our experimental resolution doesn't allow us to identify which one is the AA stripe or the AB stripe, but the map is clearly consistent with the pattern shown in Fig 1b, except with lattice relaxations. The relatively low and high conductance regions develop into stripes, with an inter-stripe distance ~ 7.1 nm, consistent with the expectation for a $\theta \sim 5°$ tWTe$_2$ stack. (4) Since WTe$_2$ is air sensitive, we have to employ a few-layer hBN as a protecting layer and a sample fabrication process that minimizes the time for the sample exposed to air. The top surface of the thin hBN is left behind with polymer residues etc, which we believe could be the main source of the residue-like features in **b** and **c**. Our transport devices (devices #1 & #2 in the main text) employ a top graphite gate that serves as a screening layer and hence the tWTe$_2$ channel is of much higher quality. Other details about the cAFM measurements can be found in the Methods.

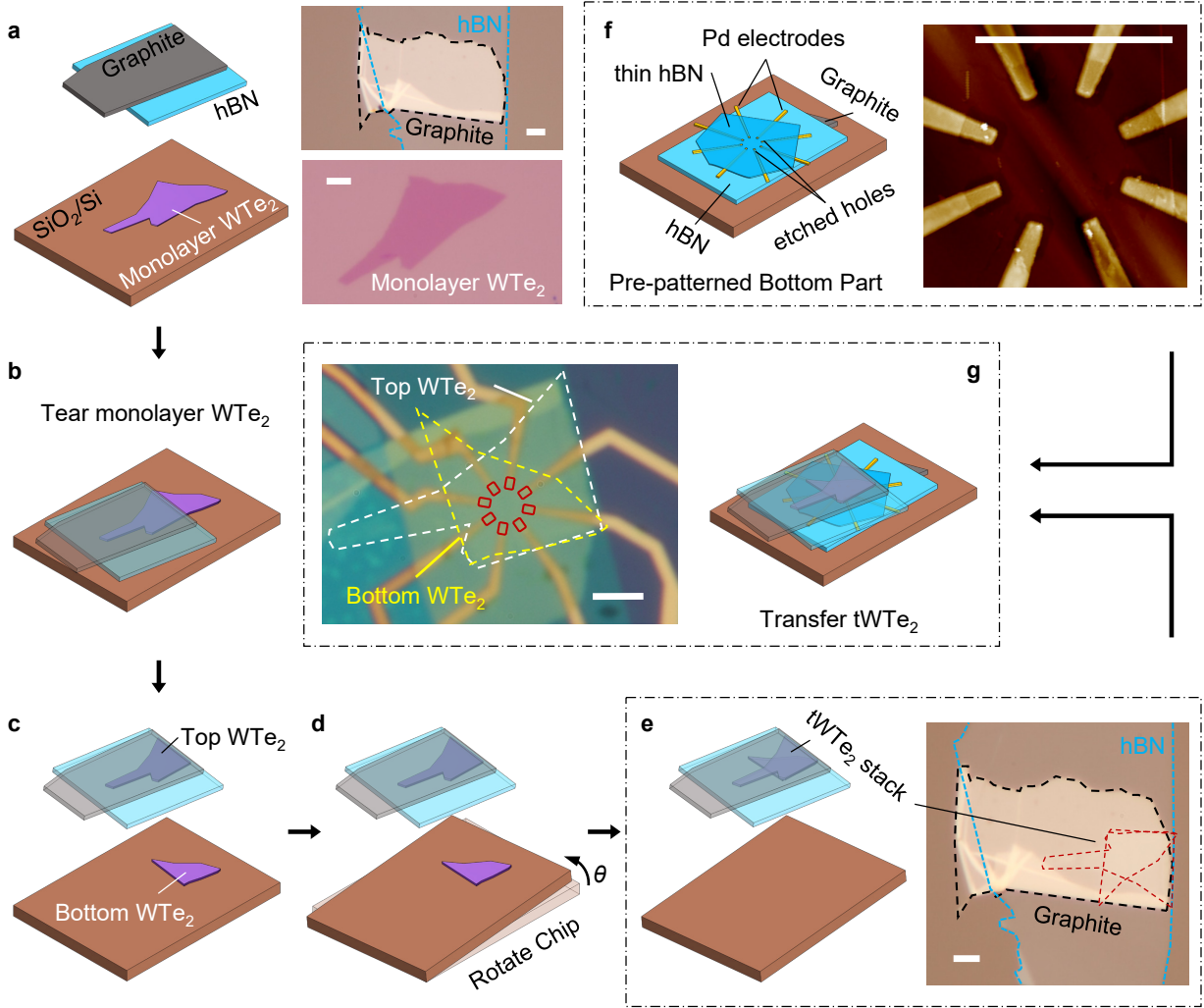

**Extended Data Fig. 2 | Sample fabrication process. a,** Cartoon illustrations of a prepared top graphite/hBN stack (top) and a flake of monolayer WTe$_2$ on a SiO$_2$/Si chip (bottom). Their corresponding optical images are also shown to the right. The distance between the aligned graphite edge and hBN edge is carefully optimized during transfer to be within 500 nm. **b & c,** Tear the monolayer WTe$_2$ by the hBN into two separate pieces, labeled as top and bottom WTe$_2$. **d,** Rotate the bottom WTe$_2$ flake, i.e., the SiO$_2$/Si chip, counterclockwise by $\theta$. **e,** Pick up the bottom WTe$_2$ to create a tWTe$_2$ stack. The optical image of the tWTe$_2$ stack shown to the right was taken after flipping the stamp upside down. The tWTe$_2$ stack is highlighted by the red dashed line. No visible bubbles were observed. **f,** The pre-patterned bottom part (thin hBN/Pd electrodes/hBN dielectric/graphite) with etched holes in the thin hBN layer to expose the tips of the Pd electrodes (see ref.[22,23]). An atomic force microscope image shows a clean surface of a prepared bottom stack. **g,** The final stack of a tWTe$_2$ device, with an optical image of a final device (#1). The thickness of flakes is 7.6 nm graphite/8.9 nm top hBN/WTe$_2$/2.0 nm thin hBN/5.5 nm bottom hBN/ 7.0 nm graphite for device #1 and 4.8 nm graphite/10.6 nm top hBN/WTe$_2$/4.2 nm thin hBN/15.0 nm bottom hBN/ 6.9 nm graphite for device #2. All scale bars are 3 μm.

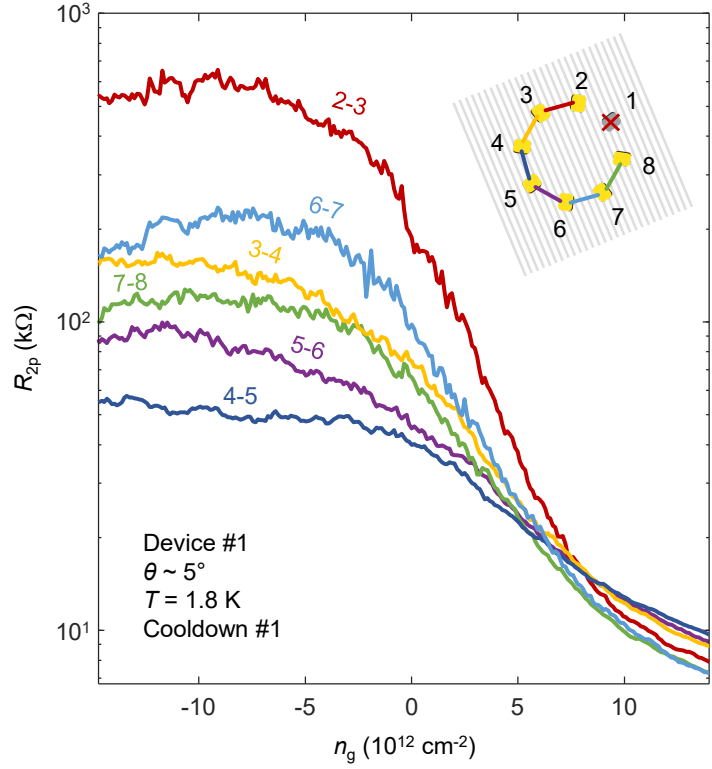

**Extended Data Fig. 3 | Two-probe resistance between neighboring electrodes as a function of $n_g$.** Inset shows the contact configurations for each measurement, where the estimated hard direction (stripe direction) is indicated by the grey lines (not to scale). $R_{2\text{-}3}$ and $R_{6\text{-}7}$ display larger values than all others in the hole-doped regime, signifying the hard direction, while $R_{4\text{-}5}$ shows the lowest value. Contact 1 was broken during the fabrication. The contact resistance plays a significant role here. After the easy and hard directions were identified, we performed four-probe measurements, as shown in Fig. 1 in the main text.

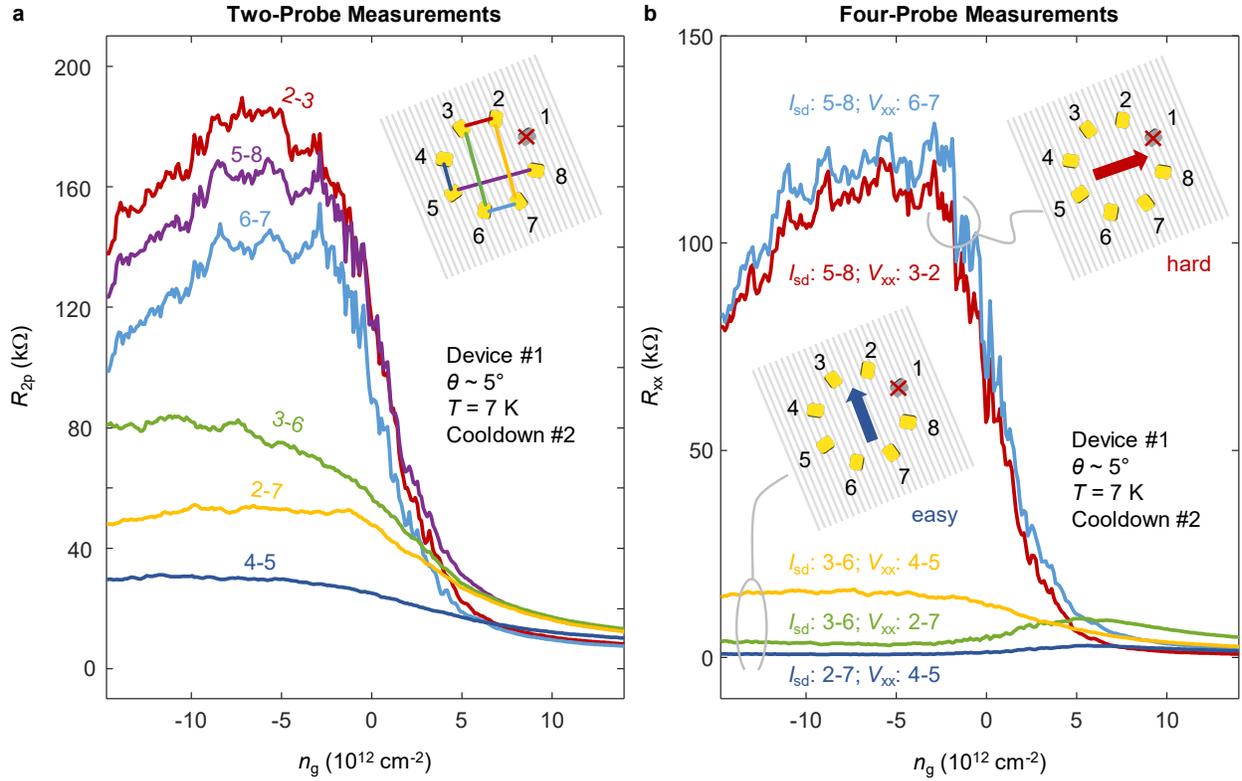

**Extended Data Fig. 4 | Two-probe and four-probe resistance across and along the stripes as a function of $n_g$ (data measured in device #1, cooldown #2). a,** Two-probe resistance across ($R_{2-3}$, $R_{6-7}$, and $R_{5-8}$) and along ($R_{4-5}$, $R_{2-7}$, and $R_{3-6}$) the wires as a function of $n_g$. Inset shows the contact configurations for each measurement, where the easy direction (along stripes) is indicated by the grey lines (not to scale). **b,** Four-probe resistance across ("$I_{sd}$: 5-8; $V_{xx}$: 6-7" and "$I_{sd}$: 5-8; $V_{xx}$: 3-2") and along ("$I_{sd}$: 3-6; $V_{xx}$: 4-5", "$I_{sd}$: 3-6; $V_{xx}$: 2-7", and "$I_{sd}$: 2-7; $V_{xx}$: 4-5") the stripes as a function of $n_g$. Both measurements further confirm the transport anisotropy in the device.

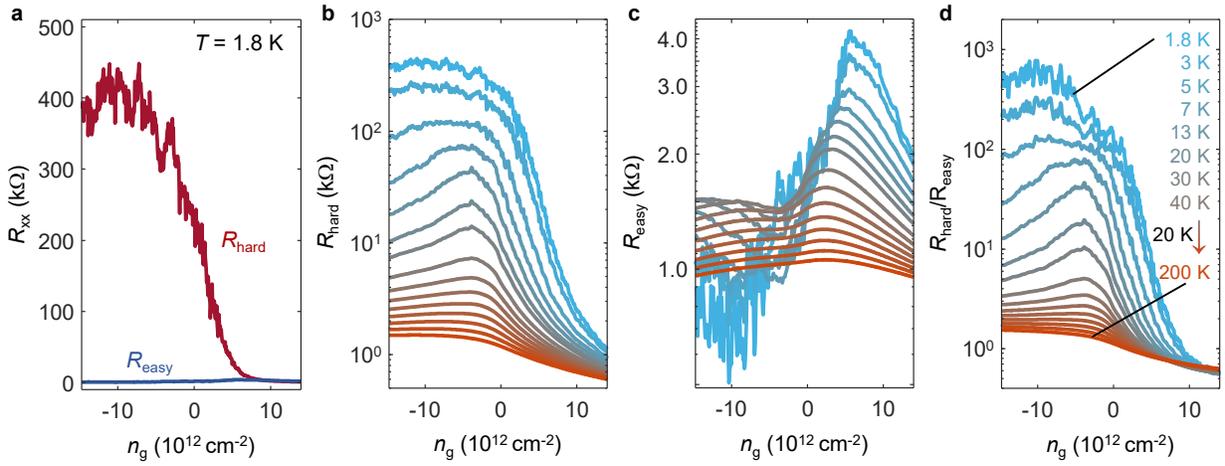
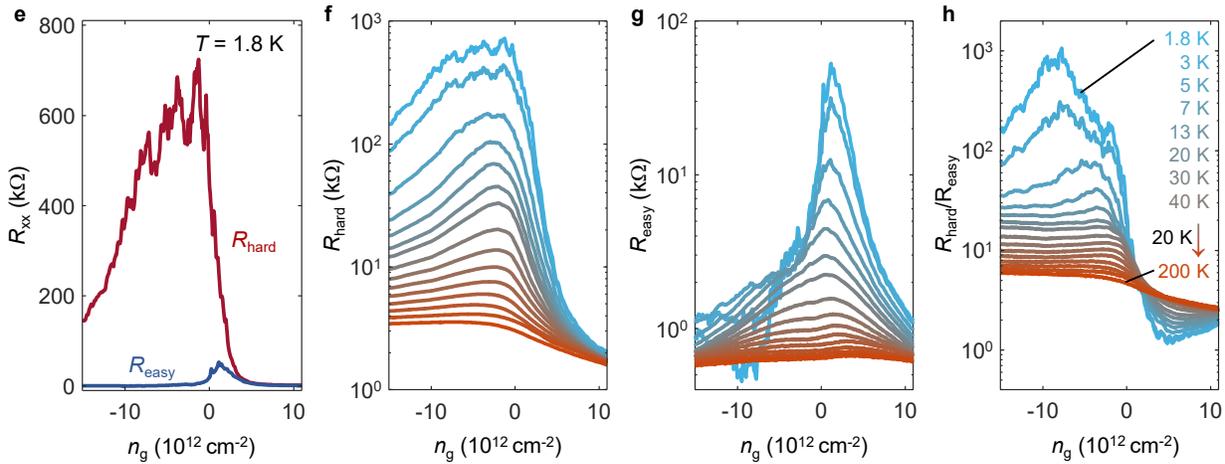

**Extended Data Fig. 5 | $n_g$ dependent transport anisotropy for devices #1 and #2.** Four-probe resistance $R_{xx}$ as a function of $n_g$ measured with an excitation current applied along hard and easy directions in linear plots. **a,** data taken for device #1 at 1.8 K (the same data as Fig. 1i). **b-d,** $R_{hard}$, $R_{easy}$, and $R_{hard}/R_{easy}$ as a function of $n_g$ at different temperatures taken from device #1. **e-h,** The same plots for device #2.

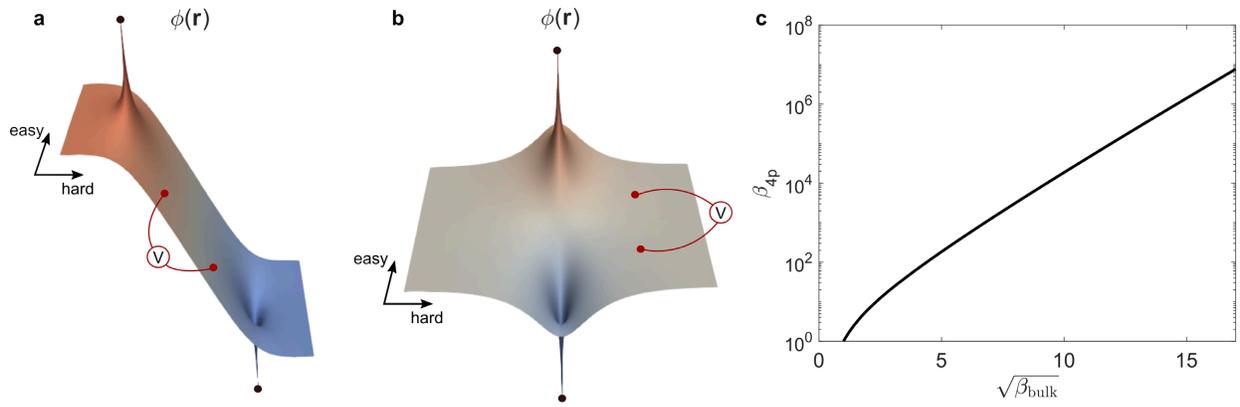

**Extended Data Fig. 6 | Electrostatic simulation for the four-probe contact configuration. a & b,** Electric potential distribution for contact arrangements corresponding to $R_{hard}$ and $R_{easy}$ four-probe measurements respectively (see Methods). Black dots indicate current contacts that source/sink current. Red dots indicate the placement of voltage contacts. **c,** Predicted four-probe anisotropy $\beta_{4p} \equiv R_{hard}/R_{easy}$ as a function of the intrinsic sheet resistivity anisotropy $\beta_{bulk}$. For $\beta_{4p} \sim 1000$, we estimate $\beta_{bulk} \sim 50$.

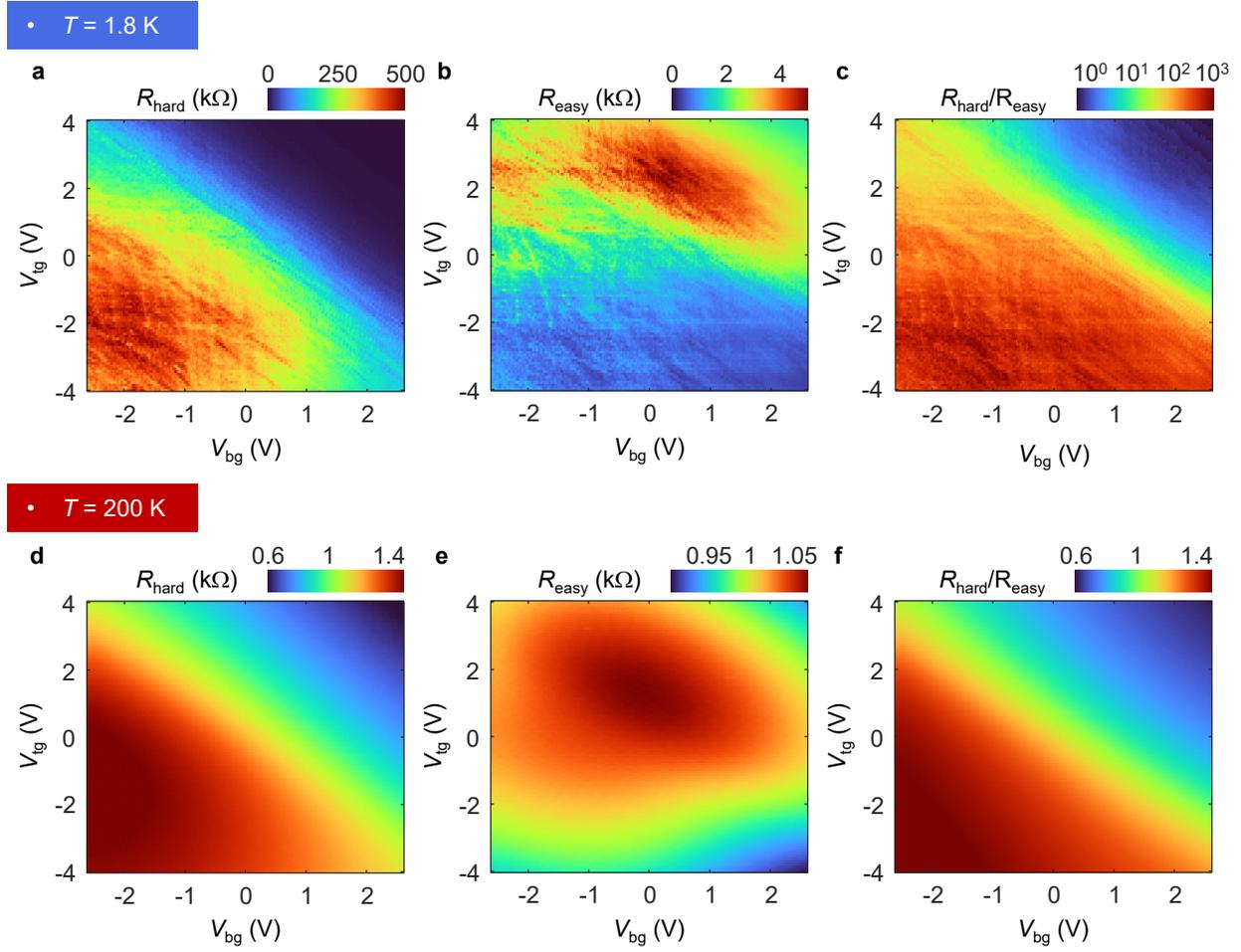

**Extended Data Fig. 7 | Dual-gate dependent transport along hard and easy directions (device #1, cooldown #1).** The four-probe resistance taken at 1.8 K (200 K) along the hard and easy directions were shown in **a** (**d**) and **b** (**e**), respectively. $R_{hard}/R_{easy}$ at 1.8 K (200 K) is shown in **c** (**f**).

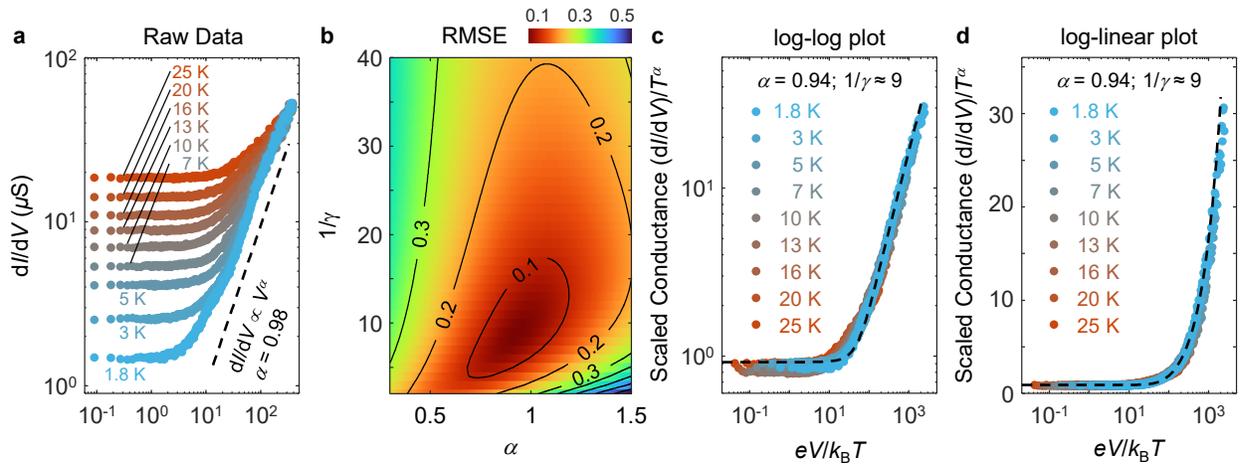

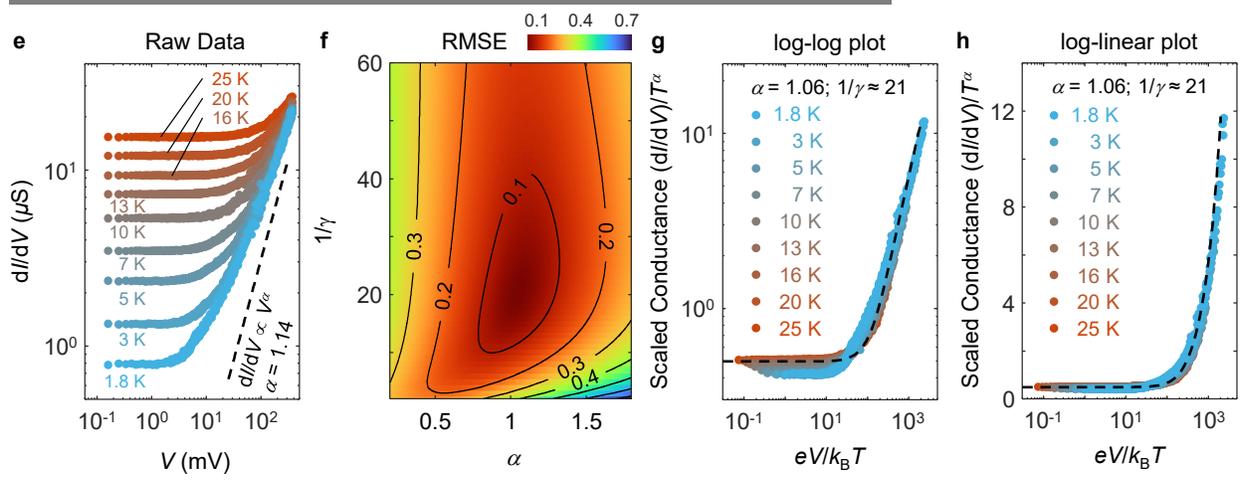

**Extended Data Fig. 8 | Fitting to the differential conductance data based on the universal scaling formula. a,** Raw data for the differential conductance measurements taken in device #1 (replotted from Fig. 2d). **b,** 2D map of calculated root-mean-square error (RMSE) as a function of the fitting parameters, $\alpha$ and $\gamma$ (see Methods for details). The best fit is obtained by finding the minimal value of RMSE in this plot, i.e., $\alpha = 0.94$ and $1/\gamma = 9$. **c & d,** Scaled conductance as a function of scaled excitation by assigning $\alpha = 0.94$ in a log-log plot and log-linear plot. The dashed line indicates the fitting result given by the universal formula defined in the Method section. **e-h**, The same fitting plots for device #2, using the same raw data shown in Fig. 2f.

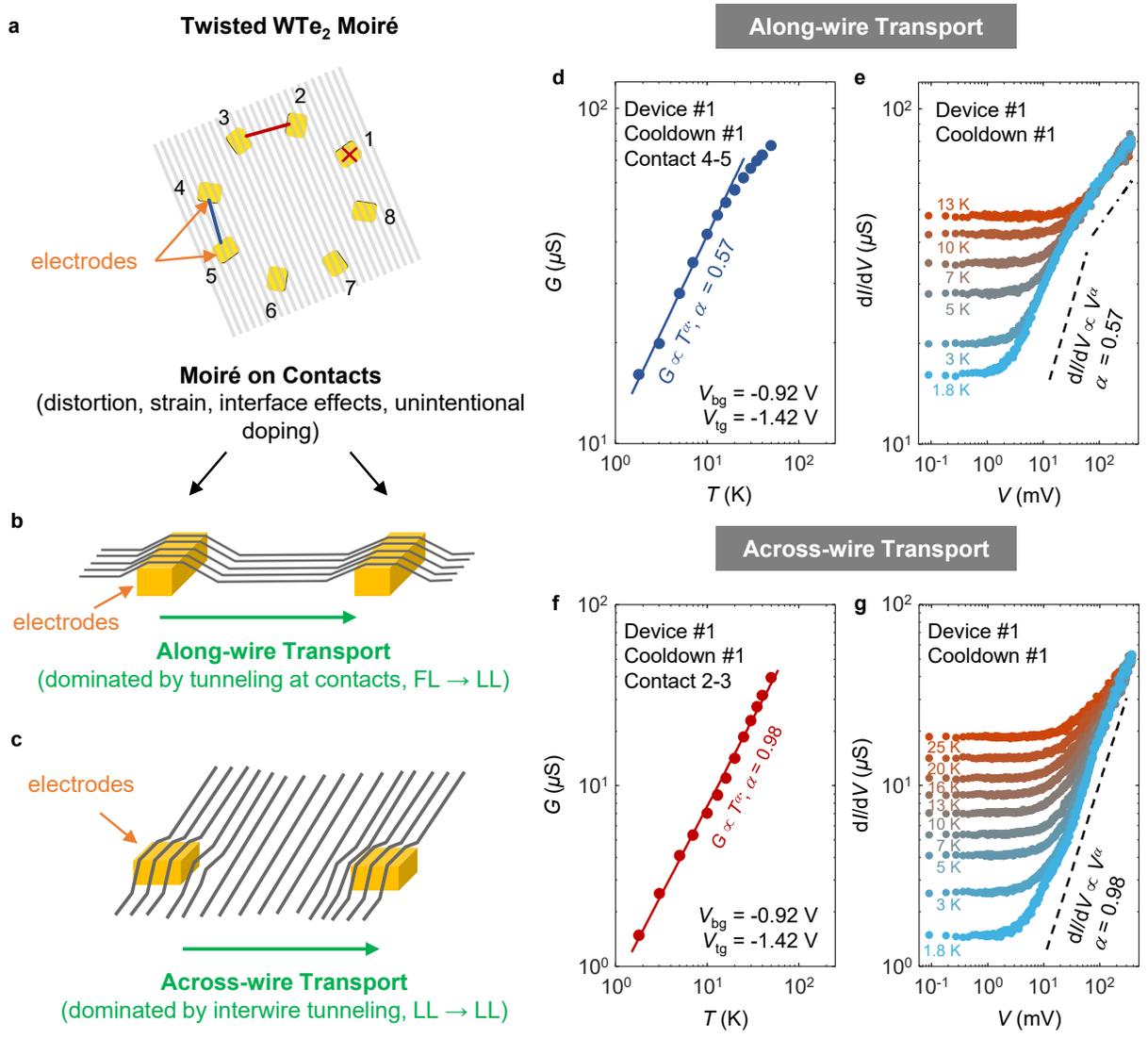

**Extended Data Fig. 9 | Comparison between along-wire and across-wire transport. a,** Illustration of tWTe$_2$ moiré stripes on the electrodes (top view). **b,** Illustration of transport along wires. At low $T$, the along-wire transport is dominated by contact resistance, i.e., tunneling from the metal (FL) to the moiré wires (LL). **c,** Illustration of the across-wire transport, where the dominant resistance is due to interwire tunneling in the stripe regime (i.e., LL to LL tunneling). **d,** Along-wire two-probe conductance $G$ as a function of $T$, plotted in log-log scale at a selected gate parameter. A power-law fit (solid line) to the low $T$ data is shown. **e,** Differential conductance d$I$/d$V$ taken under the along-wire transport configuration as a function of d.c. bias $V$ at different $T$. The dashed line indicates a power law trend. The dot-dash line indicates a deviation from the trend at high bias. Note that distortions, strain, unintentional doing, and other interface effects occur at the moiré in the contact regime, which could cause the deviation. **f & g,** the same plot for data taken from the across-wire transport (the same data as Fig. 2c and d), exhibiting a more robust power-law behavior to higher bias and $T$. This can be understood as the dominant resistance in the across-wire transport comes from the tWTe$_2$ channel regime, which is more uniform compared to the contact regime. Data were taken from device #1 in cooldown #1.
25

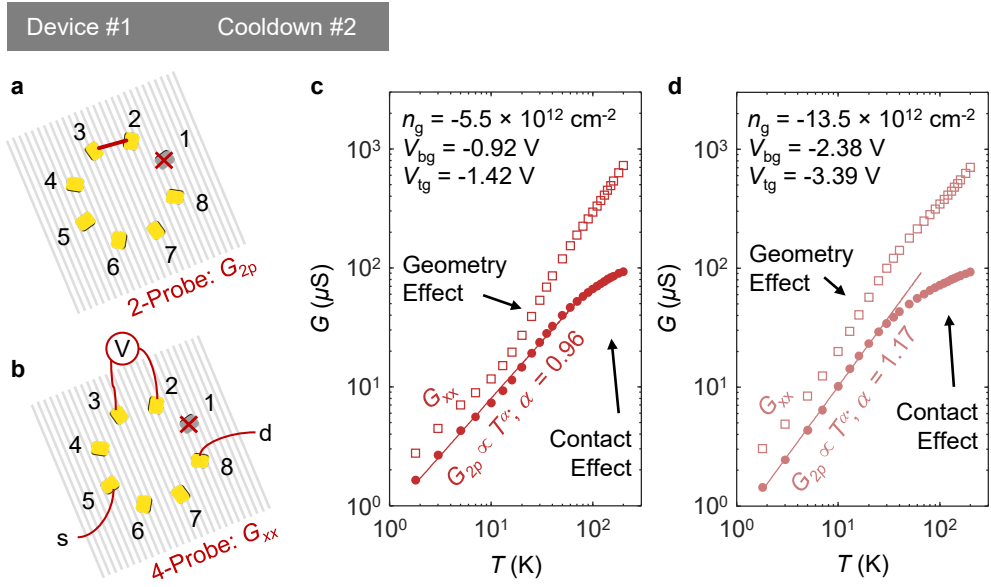

**Extended Data Fig. 10 | Comparison of two-probe and four-probe measurements across the wires.** Cartoon illustration of **(a)** two-probe ($G_{2p}$) and **(b)** four-probe ($G_{xx}$) configurations used for the measurements. **c and d,** $G_{2p}$ and $G_{xx}$ as a function of temperature taken in the hole-doped region ($n_g = -5.5 \times 10^{12}$ cm$^{-2}$ and $n_g = -13.5 \times 10^{12}$ cm$^{-2}$, respectively). At low $T$ (1.8 K ~ 25 K) the trends of $G_{2p}$ and $G_{xx}$ both follow a power law and match well, demonstrating that the power law is intrinsic to the tWTe$_2$ channel. At high $T$, the two trends of $G_{2p}$ and $G_{xx}$ deviate from each other, which can be understood as $G_{2p}$ saturates due to contact resistances while $G_{xx}$ is strongly affected by the temperature induced changes of anisotropy. The effective geometry factor, important for determining $G_{xx}$, changes as the sample is tuned from a strongly anisotropic phase at low $T$ to an isotropic phase at high $T$. The main analyses in this paper are focused on the low $T$ regime. The measurements were performed on device #1 in cooldown #2.

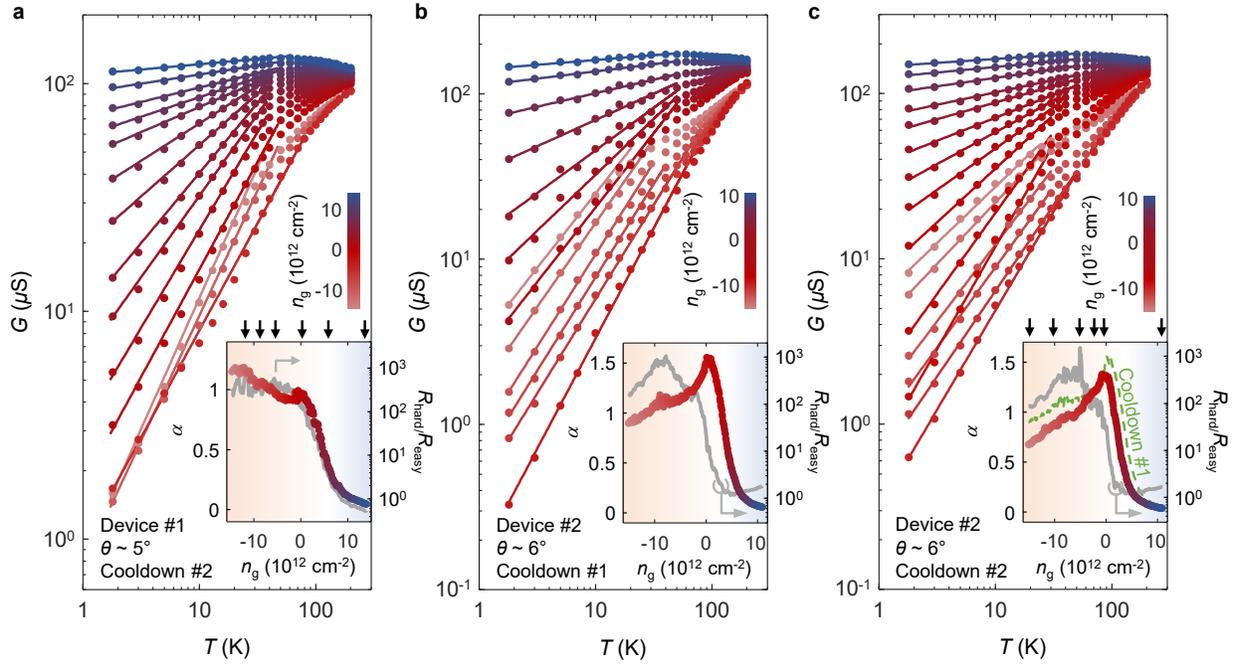

**Extended Data Fig. 11 | Gate-tuned anisotropy crossover. a,** The across-wire two probe conductance $G(T)$ displays a power-law relation ($G \propto T^\alpha$) for a wide range of doping for device #1 (cooldown #2). The color of the data points encodes $n_g$, as shown in the color bar. The solid lines are the power-law fittings, where the extracted exponent $\alpha$ is shown in the inset. The gray line replots the anisotropy ratio. **b,** The same plots for device #2 (cooldown #1). The gray line replots the anisotropy ratio shown in the inset of Fig. 1h. **c,** The same plots for device #2 (cooldown #2). Note that data taken from two different cooldowns from device #2 shows qualitatively consistent results with only minor quantitative differences (dashed line in the inset of **c** is the exponent $\alpha$ replotted from the inset of **b** for comparison). Arrows to the insets in **a** and **c** indicate the selected $n_g$, at which the scaling analysis of the differential conductance is performed in Extended Data Figs. 12 and 13, respectively.

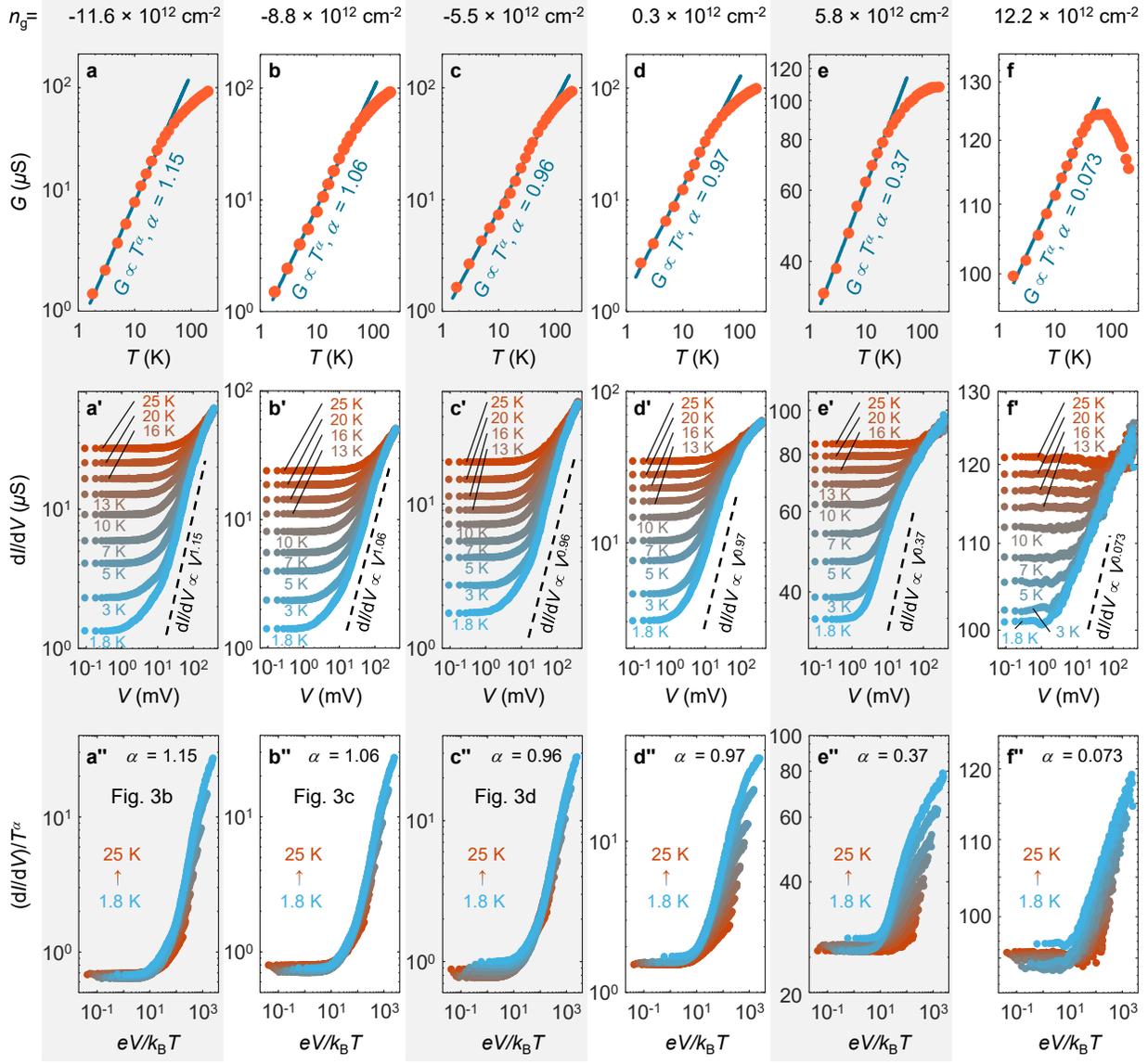

**Extended Data Fig. 12 | Additional power-law scaling analysis for device #1 (cooldown #2).** The corresponding $n_g$ for each data set is indicated in the inset of Extended Data Fig. 11a. **a,** Temperature dependent across-wire two probe conductance $G(T)$ taken at the indicated $n_g$. The solid line is the power low fit. **a',** Bias dependent differential conductance taken at the same $n_g$ under different $T$. The dashed line indicates the power-law trend with the same exponent α extracted in **a**. **a'',** the same data in **a'**, but replotted as scaled differential conductance $(dI/dV)/T^\alpha$ v.s. scaled bias $eV/k_BT$. Other panels are the same plots for different $n_g$. As seen in the plots, in the hole side (**a-c**) the data generally follows a power law very well, while near charge neutrality (**d**) and in the electron side (**e**), deviations start to develop at high bias. In the highly electron-doped region (**f**), $dI/dV$ and $G$ vary only a little bit (α ~ 0) with changing both $V$ and $T$, hence the behavior is approximately ohmic. Data used for Figs. 3b-d are indicated in the lowest panel.

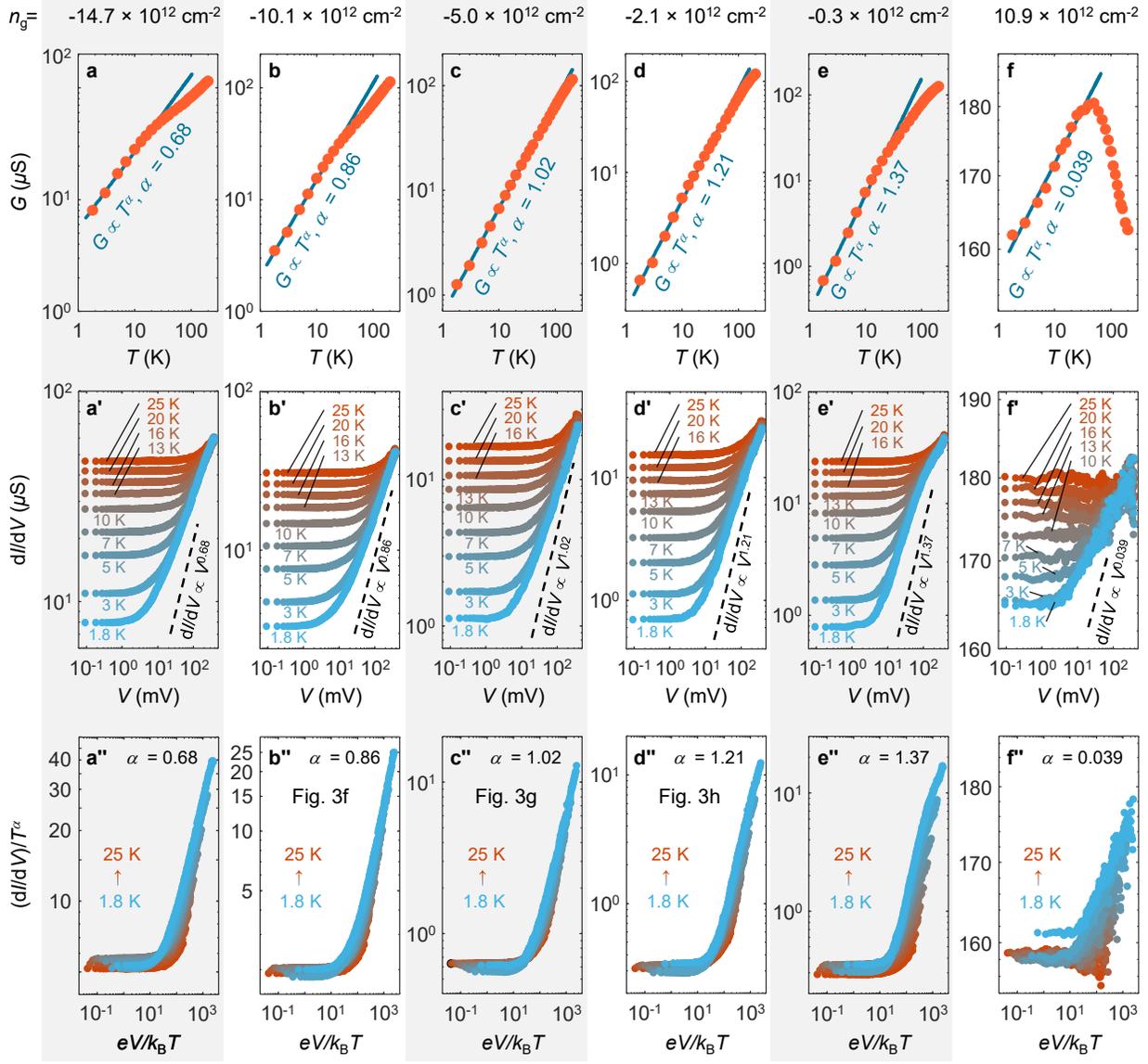

**Extended Data Fig. 13 | Additional power-law scaling analysis for device #2 (cooldown #2).** The corresponding $n_g$ for each data set is indicated in the inset of Extended Data Fig. 11c. **a,** Temperature dependent across-wire two probe conductance $G(T)$ taken at the indicated $n_g$. The solid line is the power low fit. **a',** Bias dependent differential conductance taken at the same $n_g$ under different $T$. The dashed line indicates the power-law trend with the same exponent $\alpha$ extracted in **a**. **a'',** the same data in **a'**, but replotted as scaled differential conductance $(dI/dV)/T^\alpha$ v.s. scaled bias $eV/k_BT$. Other panels are the same plots for different $n_g$. As seen in the plots, in the hole side (**a-d**) the data generally follows a power law very well, while near charge neutrality (**e**), deviations start to develop at high bias. In the highly electron-doped region (**f**), $dI/dV$ and $G$ vary only a little bit ($\alpha \sim 0$) with changing both $V$ and $T$, hence the behavior is approximately ohmic. Data used for Figs. 3f-h are indicated in the lowest panel.

29